\documentclass{article}
\usepackage{ijcai22}

\usepackage{times}
\usepackage{soul}
\usepackage{url}
\usepackage[hidelinks]{hyperref}
\usepackage[utf8]{inputenc}
\usepackage[small]{caption}
\usepackage{graphicx}
\usepackage{amsmath}
\usepackage{amsthm}
\usepackage{amssymb}
\usepackage{booktabs}
\usepackage{algorithm}
\usepackage{algorithmic}
\usepackage{enumitem}
\usepackage{forest}
\usepackage{multirow}
\title{Negative Sampling for Contrastive Representation Learning: A Review}

\author{
Lanling Xu$^{1,3}$
\and
Jianxun Lian$^{1}$\and
Wayne Xin Zhao$^{4}$\footnote{Corresponding author}\and 
Ming Gong$^2$\and \\
Linjun Shou$^2$\and
Daxin Jiang$^2$\and
Xing Xie$^1$\And
Ji-Rong Wen$^{3,4}$
\affiliations
$^1$Microsoft Research Asia\and
$^2$Microsoft STC Asia\\
$^3$School of Information, Renmin University of China\\
$^4$Gaoling School of Artificial Intelligence, Renmin University of China
\emails
xulanling\_sherry@163.com,
\{jianxun.lian\}@microsoft.com,
batmanfly@gmail.com, \\
\{migon, lisho, djiang, xingx\}@microsoft.com,
jrwen@ruc.edu.cn
}

\begin{document}

\maketitle


\begin{abstract}
The ``learn-to-compare'' paradigm of contrastive representation learning (CRL), which compares positive samples with negative ones for representation learning, has achieved great success in a wide range of domains, including natural language processing, computer vision, information retrieval and graph learning. While many research works focus on data augmentations, nonlinear transformations or other certain parts of CRL, the importance of negative sample selection is usually overlooked in literature. In this paper, we provide a systematic review of negative sampling (NS) techniques and discuss how they contribute to the success of CRL. As the core part of this paper, we summarize the existing NS methods into four categories with pros and cons in each genre, and further conclude with several open research questions as future directions. By generalizing and aligning the fundamental NS ideas across multiple domains, we hope this survey can accelerate cross-domain knowledge sharing and motivate future researches for better CRL.
\end{abstract}

\section{Introduction}

In deep learning, to make a good representation, a series of methods collectively referred to as contrastive learning (CL) provide a simple yet effective ``learn-to-compare'' paradigm. Despite its success in recent self-supervised learning \cite{SSL}, contrastive learning is not a new idea, with related works stretching back for over a decade \cite{CL-early}, ranging from supervised, semi-supervised to self-supervised methods in texts, images, retrievals, graphs, etc. Generally speaking, we consider all methods that contrast between positive and negative pairs to learn representations as contrastive representation learning (CRL) \cite{CRL}. 




The induced representations of CRL consist of two key properties: \textit{alignment} of features from positive pairs and \textit{uniformity} of representation on the hypersphere \cite{uniformity-alignment}, and the latter mainly depends on separation of negative pairs. That is to say, solely minimizing the distance between positive samples will make the distance between any samples close to zero, and the quality of negative pairs has a decisive impact on the representation from catastrophic collapse. Due to expensive manual labeling and the lack of explicit negative signals in most cases, the strategy of negative sampling (NS) is to select (or generate) negative samples for a given positive pair from negative candidates. 

\noindent \textbf{Motivation: why is negative sampling important?}

As described above, ``negative'' samples are introduced to keep the \textit{uniformity} property to avoid representation collapse \cite{SimCSE}, while ``sampling'' is a useful and practical strategy from both efficient and effective perspective. On the one hand, sampling greatly improves the computational efficiency. Specifically, the number of candidate negatives is generally equal to or even far more than the whole data set (e.g., over 2 billion candidates for each pair \cite{PinSage}). Therefore, the computational burden of comparing each positive pair with all negative pairs is too heavy to accept with limited computing power, which may also submerge the gradient of positive pairs against too many negative samples \cite{RCNN}. On the other hand, sampling informative negatives further promotes the training results. In objective optimization, it is the informative negative that plays a crucial role, while most negatives are easy to discriminate with minor gradient contributions and a few negatives are even counter-productive \cite{negatives-analysis}. Regardless of efficiency, taking all candidates equally as negative samples is not the optimal solution in most cases.

Owing to success of CRL, steady progress has been made in developing algorithms for data augmentations \cite{SimCLR}, nonlinear transformations \cite{NGCF} or other certain parts of CRL. However, the choice of negative samples gets less attention, which is usually ignored or simplified by random selection \cite{CL-early}. Since the naive strategy is independent with model parameters, easy negatives are chosen by random sampling with high probability, resulting in slow convergence and significant performance loss. For instance, the convergence rate of an adaptive sampling strategy for item recommendation is more than 10 times faster than that of random selection with better results \cite{AdaBPR}, and sometimes the results can be even improved by nearly 50\% for entity alignment \cite{truncated-KGE}. Numerous empirical results have shown that CRL benefits from more intelligent negative sampling strategies.

\noindent \textbf{Concentration: what makes a good negative sampler?}

Given the important role of negative sampling for CRL, we review a comprehensive set of related methods across various fields, summarize their advantages and disadvantages, and conclude that the core properties for negative samplers can be evaluated in four dimensions.

\begin{itemize}[leftmargin=*]
    \item \textit{Efficient.} The negative sampler should have low time and space complexity, which hardly affects overall resources.
    \item \textit{Effective.} The informative samples from the negative sampler should lead the model towards a constringent acceleration and gradient change on the aspect of rationality.
    \item \textit{Stable.} The negative sampler shall be robust and data-insensitive with little probability to collapse.
    \item \textit{Data-independent.} The negative sampler should not rely heavily on the side information. Otherwise, data availability will greatly restrict the practical applicability.
\end{itemize}

At present, there is no method that meets all the above requirements, raising the question of trade-offs among various properties in employing negative sampling for CRL.

\noindent \textbf{Contributions}. Due to the prevalence of contrastive learning in self-supervised representation learning (SSL), there exist several comprehensive surveys on CL\cite{CSSL}, SSL\cite{SSL} or CRL\cite{CRL}. However, none of them concentrate on negative sampling, which is a must for generalized CRL based on ``learn-to-compare'' paradigm. To the best of our knowledge, this paper is the first review focusing on negative sampling methods for CRL, and the main contributions of this work are summarized below:
\begin{itemize}[leftmargin=*]
    \item We formulate the general task of negative sampling for CRL in texts, images, retrievals and graphs, providing a up-to-date review and compendious sum-up. Through this article, one can easily grasp the development history and frontier ideas of negative sampling across multiple AI fields. 
    \item We divide negative sampling methods into four categories according to distinguishing features, and the advantages and disadvantages of each one are discussed systematically.
    \item We conclude several open research problems in this field, in order to present a pointer to relevant researches on the application of negative sampling for but not limited to CRL.
\end{itemize}

\begin{table*}
\small
\begin{center}
\begin{tabular}{@{}l|l|l@{}}
\toprule
Category                                       & Subcategory                          & Representative Approaches    \\ \midrule
\multirow{2}{*}{Static NS}      & Random NS             & Random selection \cite{CL-early}             \\ \cmidrule(l){2-3} 
                                               & Popularity-biased NS  & Word2vec \cite{word2vec}; PNS \cite{AdaBPR}          \\ \midrule
\multirow{6}{*}{Dynamic NS}  & Query-dependent DNS                & 
  \begin{tabular}[c]{@{}l}DNS \cite{DNS} \cite{NLP-DNS-2018}; ANCE \cite{ANCE}\\  PinSage \cite{PinSage}; IS \cite{unsupervised-importance-sampling} \end{tabular} \\ \cmidrule(l){2-3} 
                                               & Positive-dependent DNS & 
  \begin{tabular}[c]{@{}l}Max-sampling \cite{max-sampling}; GNEG \cite{GNEG}  \\  $\epsilon-$Truncated UNS \cite{truncated-KGE}; TransEdge \cite{TransEdge} \end{tabular} \\ 
\cmidrule(l){2-3} & Hybrid DNS & 
  \begin{tabular}[c]{@{}l}FaceNet \cite{FaceNet}; WAPR \cite{warp-2010} \\ MoChi \cite{MoChi}; MixGCF \cite{MixGCF} \end{tabular} \\ \midrule
\multirow{4}{*}{Adversarial NS} & Discrete Sampling           & 
  \begin{tabular}[c]{@{}l}IRGAN \cite{IRGAN}; GraphGAN \cite{GraphGAN} \\ KBGAN \cite{KBGAN}; KGPolicy \cite{KGPolicy} \end{tabular} \\ \cmidrule(l){2-3} 
& Continuous Sampling         & 
  \begin{tabular}[c]{@{}l}DAML \cite{DAML}; AdvIR \cite{AdvIR} \\ HeGAN \cite{HeGAN}; AdCo \cite{AdCo} \end{tabular} \\ \midrule
\multirow{2}{*}{Efficient NS} &
  In-Batch NS &
  \begin{tabular}[c]{@{}l}SimCLR \cite{SimCLR}; PBG \cite{PBG} \\ ABS \cite{ABS}; GraphVite \cite{Graphvite} \end{tabular} \\ \cmidrule(l){2-3} 
                                               & Caching Mechanism      & NSCaching \cite{NSCaching}; MoCo \cite{MoCo}              \\ \bottomrule
\end{tabular}
\end{center}
\vspace{-0.15in}
\caption{Categorization of negative sampling approaches for CRL.}\label{table1}
\end{table*}

\section{Task Formulation and Typical Applications}

The ultimate goal of CRL is to learn distinguishable information to pull together positive pairs (\textit{similar} ones), and push apart negative pairs (\textit{dissimilar} ones). {To contrast between positive and negative pairs, InfoNCE \cite{InfoNCE} is widely accepted as a generalized form of other comparison-based losses \cite{SSL}.}
Without loss of generality, we formulate negative sampling for CRL as follows:


\begin{align*}
    \mathcal{L(\theta)} &= \mathbb{E}_{p(x, y^+, y^-)}l_{\theta}(x, y^+, y^-) \\
    &= \mathbb{E}_{\mathop{}_{\boldsymbol{y^-\sim p(y^-)}}^{(x, y^+)\in \mathcal{T}}}[-{\rm log}(\dfrac{e^{\phi_{\theta}(x, y^+)}}{e^{\phi_{\theta}(x, y^+)} + \sum_{k=1}^K e^{\phi_{\theta}(x, y^{-}_{(k)})}})]
\end{align*}
where $x$ is the anchor or query, $y^+$ is the positive sample and $y^-$ denotes the negative sample. $\mathbb{E}_{p(x, y^+, y^-)}$ is the expectation with respect to the joint distribution $p(x, y^+, y^-)$ of samples. $\phi_{\theta}(\cdot)$ is used to measure the score of samples, such as distance and similarity with parameters $\theta$ in the latent space. $l(\cdot)$ captures the loss that scores a positive tuple $(x, y^+)$ against $K$ negative ones in the form of InfoNCE. 

Focusing on negative sampling, despite the variation in form, the essence is to learn the distribution of negative samples (i.e., $y^-\sim p(y^-)$) for contrastive comparison, and $\sim$ denotes sampling from the specific distribution. As is classified in Section~\ref{categorization}, different kinds of negative distributions correspond to different negative sampling approaches for CRL. 


{In general, the existing CRL methods involving negative sampling can all be reduced into our formulation, and the target set $\mathcal{T}$ vary from task to task. As follows, we introduce four representative applications in different fields.}

\begin{itemize}[leftmargin=*]
    \item In natural language processing (NLP), typical CRL tasks include word embedding \cite{word2vec} and sentence embedding \cite{SimCSE}. $\mathcal{T}=\{(\omega_I, \omega_O)\}$ where $\omega_I$ means the input such as context words, and $\omega_O$ is the output such as the target word. Given $\omega_I$, negative sample $\omega_i$ is selected from corpus based on certain strategies, i.e., negative sampling.
    \item In computer vision (CV), CRL has been applied in tasks such as image classification \cite{SimCLR}, $\mathcal{T}=\{(q, p)\}$ where $q$ and $p$ denote two augmented views of the same image, or the query image and the related one. Without supervision signals, negative samples are mainly utilized for pushing apart views from different images, while label information can be leveraged in a supervised setting.
    \item In information retrieval (IR), $\mathcal{T}=\{(q, d)\}$ where $q$ is the query and $d$ is the relevant document. In specific tasks such as implicit collaborative filtering (ICF) \cite{DNS}, $\mathcal{T}=\{(u, i)|u\in \mathcal{U}, i\in \mathcal{I}\}$, where $u$ and $i$ denotes the user and item of an interaction. $\mathcal{U}$ and $\mathcal{I}$ are user and item sets. With pairwise inputs, negative sampling aims at selecting negative samples for users from non-interactive items. 
    \item In graph representation learning (GRL), the encoded node representation can be learned in a similar way as word embeddings \cite{LINE}. Particularly in knowledge graph embedding (KGE) tasks \cite{truncated-KGE}, $\mathcal{T}=\{(h,r,t)|h,t\in \mathcal{E}, r\in \mathcal{R}\}$. $h$, $r$ and $t$ denotes the head entity, relation and tail entity of a triple. The corrupted false triples are constructed by replacing the head or tail entity with a certain negative sampling strategy. 
\end{itemize}




\section{Negative Sampling Approaches for CRL}
\label{categorization}

In this section, we systematically review representative negative sampling methods from different domains, and divide them into four categories, with particular genres in each one, for better understanding and comparisons. The general categorization is illustrated in Table~\ref{table1}. 

\subsection{Static Negative Sampling}

As is shown in Algorithm~\ref{alg:algorithm-gsns}, static negative sampling assigns a static, global probability for each candidate to be sampled, where $\mathcal{E}_x$ denotes positive samples of $x$ and $\mathcal{E}_{y^-}$ denotes the sampled negatives, the same below. Typically, the uniform and popularity-biased distribution correspond to the random negative sampling (RNS) and popularity-biased negative sampling (PNS) respectively. To make it more formal, $p(y^-)\propto deg(y^-)^{\alpha}$, namely, the probability selected as the negative is proportional to $\alpha$ power of $y^-$'s popularity value (such as word frequency in word embedding). RNS can be regarded as a simple case of PNS at $\alpha=0$.

\paragraph{Random Negative Sampling.} RNS is the most basic negative sampling method, which is often used as the default sampling strategy and basic baseline for its simplicity and plainness. It is generally believed that a strategy which cannot stably outperforms RNS is of no practical use. 
In a broad sense, a random negative selection
can all be regarded as the applications of RNS though it is not explicitly stated. 

\paragraph{Popularity-biased Negative Sampling.} 
\cite{OCCF} introduces user-oriented and item-oriented sampling to adjust the negative distribution according to the number of interactions in one-class collaborative filtering, which is an early attempt to consider data frequency in negative sampling. In the later recommendation systems \cite{AdaBPR}, PNS usually takes the popularity of items as the unigram negative distribution, that is, it tends to choose popular items as the negative samples. 
Negative sampling in natural language processing originates from word2vec \cite{word2vec}, which finds that the popularity-biased distribution at $\alpha=3/4$ (i.e., $P_n(w)=U(w)^{3/4}/\Sigma_i^{|vocab|}U(w_i)^{3/4}$) significantly outperforms the unigram and uniform distributions according to experimental experience of authors. The training paradigm and hyper parameter of word2vec is also used in tasks such as graph learning \cite{LINE} and word embeddings \cite{GNEG}.  Despite its wide application, $\alpha=3/4$ is an empirical value that lacks theoretical explanation,  the best results can even be achieved when $\alpha$ is negative in \cite{word2vec-applied-recommendation}, indicating the uncertainty of popularity on the quality of negative samples. 

\paragraph{Pros and cons.} 
Static NS is the most simple approach for implementation. It is stable, fast, and auxiliary data-independent. As formulated by Oord \textit{et al} \shortcite{InfoNCE}, more negative samples lead to a tighter lower bound on mutual information. When the number of negative samples is large enough, static NS can achieve satisfying performance. However, in many cases such as pre-training large NLP models, only a small number of negative samples can be afforded during training with GPUs. In addition, the static setting of candidate probability distribution means that negative samples do not change dynamically in pace with the training process or adaptively adjust the negative distribution tailored to each query, resulting in low-quality negative samples along with sub-optimal data representations.

\subsection{Dynamic Negative Sampling} 
\label{sec:DNS}

Due to the inherent limitations of easy negative samples, quite a few researchers have studied how to mine hard negatives which are difficult to distinguish from the positive samples. Related works can be traced back to \cite{bootstrap-1998}, which adds the images misjudged by the current classifier as negative samples to expand the negative sample set so as to improve model performance.  This bootstrapping-like approach lays foundations for subsequent dynamic negative sampling (DNS) methods for CRL, such as hard example mining (HEM) \cite{OHEM-bootstrap} in computer vision.

In general, DNS is an extension of bootstrapping, it dynamically selects contrastive pairs that are difficult to discriminate for the current model. In this way, the samples confusing the model can be highlighted in the training process, so that model parameters can be fully optimized. Formally, the selection of negative samples is determined by candidates' propensity scores, which reflect the quality of samples and will vary with the training process: $y^-\sim p(y^-|(x, y^+))\propto f(x, y^+, y^-)$, where $f(\cdot)$ is a scorer that produces a propensity score for $y^-$. Specifically, it is a mutually reinforcing feedback mechanism that negative sampler selects high-quality contrastive pairs for the model, and the gradually optimized model provides guidance for distribution of candidates. The dynamic negative sampling framework can be generalized as in Algorithm~\ref{alg:algorithm-gdns}. According to the propensity function of DNS, relevant methods can be roughly divided into \textit{Query-dependent DNS}, \textit{Positive-dependent DNS} and \textit{Hybrid DNS}.

\paragraph{Query-dependent DNS.} The propensity score is based on the predictive score of the pair \textit{$\langle$ query, sample candidate $\rangle$}. A higher score implies the current model mis-classifies the candidate sample as a positive one, so this candidate will be selected as the negative with a greater probability, i.e., $y^-\sim p(y^-|x)\propto {\phi}_{\theta}(x, y^-)$. One of the representative methods is dynamic negative sampling (DNS) \cite{DNS}, which selects hard samples with higher predicted scores during each update and significantly outperforms RNS, revealing the importance of adaptive sampling methods according to model parameters. The idea of DNS can also be applied to other tasks including but not limited to word representation \cite{NLP-DNS-2018} and dense text retrieval \cite{ANCE}. For simplicity, some methods further utilize heuristic measures to calculate weights, such as PageRank score in PinSage \cite{PinSage}. Without supervision, the distribution of hard negative pairs whose embeddings are similar to the query, can also be approximated by importance sampling (IS) strategy for SSL, and one can control the hardness of negative samples by hyperparameters \cite{unsupervised-importance-sampling}. 

\paragraph{Positive-dependent DNS.} Utilizing model scoring is to select more difficult negative samples from the perspective of objective optimization, while positive-dependent DNS makes intuitive use of the similarity between candidates and the positive sample, i.e., $y^-\sim p(y^-|y^+)$. For example, max sampling \cite{max-sampling} selects the most relevant negative answers by maximizing their similarities to the positive one for answer selection. Sun \textit{et al}. propose a similar approach named $\epsilon$-Truncated UNS \cite{truncated-KGE} to choose positive entity's nearest neighbors in the embedding space as candidate negatives, and this strategy is commonly adopted in subsequent KGE models such as TransEdge \cite{TransEdge}. Besides the induced representation in the embedding space, the semantic and structural information of existing or constructed graphs can also be employed for similarity measures. As the improvement of negative sampling in word2vec, GNEG \cite{GNEG} first constructs a co-occurrence network according to the semantic concurrence information of words in the corpus, and then obtains hard negative samples through random walk on the positive node. 

\paragraph{Hybrid DNS.} Combined with the two above, queries and positive samples can both be utilized for DNS at the same time, which we call Hybrid DNS. In this case, a proposal set of negative samples are generated by a query-dependent DNS, then positive samples will be used to filter out inappropriate candidates or incorporate new hard candidates. On the one hand, the predicted scores of positive pair can be employed as the standard to avoid hardest negatives. For instance, FaceNet \cite{FaceNet} selects \textit{semi-hard} negative samples nearest to the query in the embedding space while they are still farther than the positive-query pairs. On the other hand, positive samples can be further utilized to mine harder negatives, such as uniform sampling until a rejection in WARP \cite{warp-2010}. Considering both hard negatives and positive samples, another promising line of research involves hard negative synthesis in the embedding space as needed. Motivated by success of data mixing techniques in images, MoChi \cite{MoChi} synthesizes harder negative samples by mixing existing Query-dependent hard negative embeddings and the positive embedding directly in the latent space. Similar to MoChi's mechanism, MixGCF \cite{MixGCF} makes use of the neighborhood aggregation in graph neural network and designs positive mixing and hop mixing strategies. The former incorporates the positive embedding with candidate negatives, while the latter generates the synthetic negative embedding by combining hardest hops of the samples mixed before. 

\begin{algorithm}[tb]
\caption{A generalized framework for static NS}
\label{alg:algorithm-gsns}
\textbf{Input}: $x$, $\mathcal{E}_x$, $p(y^-)\propto deg(y^-)^{\alpha}$ \\
\textbf{Output}: $\mathcal{E}_{y^-}$
\begin{algorithmic}[1] 
\STATE $\mathcal{E}_{y^-} \leftarrow \emptyset$; 
\FOR{$k$ $\leftarrow$ $1$ to $K$}
\STATE sample $y^k \sim p(y^-)$ from $\mathcal{E} \setminus \mathcal{E}_x$;
\STATE insert($\mathcal{E}_{y^-}, y^k$);
\ENDFOR
\end{algorithmic}
\end{algorithm}

\begin{algorithm}[tb]
\caption{A generalized framework for hard NS}
\label{alg:algorithm-gdns}
\textbf{Input}: $x$, $y^+$, $\mathcal{E}_x$ \\
\textbf{Output}: $\mathcal{E}_{y^-}$
\begin{algorithmic}[1] 
\STATE{Uniformly sample $Neg$ from $\mathcal{E} \setminus \mathcal{E}_x$;}
\STATE{$\mathcal{E}_{y^-} \leftarrow \emptyset$;}
\FOR{$l$ $\leftarrow$ $1$ to length($Neg$)}
\STATE{$score$ $\leftarrow f(x, y^+, y^l)$;}
\ENDFOR
\STATE{$p(y^-|(x, y^+))$ $\leftarrow$ normalize($score$);}
\FOR{$k$ $\leftarrow$ $1$ to $K$}
\IF{Sampling}
\STATE{Sample $y^k \sim p(y^-|(x, y^+))$ from $Neg$;}
\ELSIF{Synthesis}
\STATE{Mix several $y^-\sim p(y^-|(x, y^+))$ from $Neg$ and combine with $y^+$ to synthesize $y^k$ in latent space;}
\ENDIF
\STATE{insert($\mathcal{E}_{y^-}, y^k$);}
\ENDFOR
\end{algorithmic}
\end{algorithm}

\paragraph{Pros and cons.} As the opposite of easy negatives that have already been discriminated as the negative, those easily confused hard negative samples bring greater gradient contribution to the model optimization and accelerate the convergence \cite{unsupervised-importance-sampling}. Compared with static ways, relevant studies in multiple fields have shown the effectiveness of DNS for CRL. 

However, 
although DNS can accelerate the convergence speed during training, it takes more time to calculate or update propensity scores. 
More importantly, 
the hardness of DNS is determined by hyperparameters such as the size of candidate sampling pool, and hard samples are more likely to become false negative samples (i.e., positive ones in the future data) than easy negatives. Therefore, achieving stable results of DNS has been a matter of trial and error. Actually, hardness is not the aim but the means to get informative samples, while false negatives are difficult enough with information loss. To alleviate the false negative problem, DNS usually adopts an indirect way to avoid the hardest negatives, such as empirically selecting samples whose scores are within the predefined range \cite{PinSage}, but the problem remains unresolved. Due to the existence of false negatives, blindly seeking for harder negative samples may even degrade the performance \cite{negatives-analysis}, which is the biggest challenge and limitation of DNS. 

\begin{algorithm}[tb]
\caption{A generalized framework for adversarial NS}
\label{alg:algorithm-gans}
\textbf{Input}: $\mathcal{T}=\{(x, y^+)\}$, $G$, $D$ \\
\textbf{Output}: $\theta_G$, $\theta_D$
\begin{algorithmic}[1] 
\STATE Initialize or Pretrain $\theta_G$ and $\theta_D$ for $G$ and $D$;
\REPEAT 
\FOR {Discriminator $D$ training}
\STATE Sample a mini-batch $\mathcal{T}_{batch}$ from $\mathcal{T}$;
\STATE Generate $y^-\sim p_{G}$ for each query $x$ in $\mathcal{T}_{batch}$;
\STATE Update $\theta_D$ to discriminate samples;
\ENDFOR
\FOR {Generator $G$ training}
\STATE Sample a mini-batch $\mathcal{T}_{batch}$ from $\mathcal{T}$;
\STATE Generate $y^-\sim p_{G}$ for each query $x$ in $\mathcal{T}_{batch}$;
\STATE Calculate accuracy $r$ of $D$ on sampled data;
\STATE Update $\theta_G$ to minimize the accuracy $r$;
\ENDFOR
\UNTIL{convergence} 
\end{algorithmic}
\end{algorithm}

\subsection{Adversarial Negative Sampling} 

Inspired by the popular generative adversarial networks (GAN) \cite{GAN} in recent years, researchers have proposed adversarial negative sampling methods in multiple domains. The training process of GAN can be viewed as two players playing a minimax game: one produces fake positive samples (i.e., the generator), while the other tries to discriminate them from real positive samples (i.e., the discriminator). The optimization target for the minimax game can be formalized as follows:

\begin{equation*}
\resizebox{.99\linewidth}{!}{$
    \mathop{\rm min} \limits_G \mathop{\rm max} \limits_D (\mathbb{E}_{p_{data}(y^+)}[{\rm log}D(y^+)] + \mathbb{E}_{p_{G}(y^-)}[{\rm log}(1 - D(G(y^-))])
$}
\end{equation*}
where $G$ and $D$ denote the generator and the discriminator respectively, $p_{data}$ is the real data distribution and $p_{G}$ is the pseudo distribution generated by $G$. 

When GAN is applied to negative sampling, 
the generator usually serves as a sampler to provide high-quality negative samples for confusing the discriminator, and adaptively appropriates the underlying distribution of negative samples with an alternating training procedure. In essence, dynamic negative sampling can be viewed as a simplified form of adversarial negative sampling, in which the target model itself plays as the generator to provide the distribution of negatives. Therefore, the main goal of adversarial negative sampling is still to generate hard negative samples, and the adversarial negative sampling can be generalized in Algorithm~\ref{alg:algorithm-gans}. Based on GAN, the related methods can be roughly divided into two categories: 1) to select from existing samples by the generated distribution (i.e., discrete sampling strategy), and 2) to generate new samples from the continuous embedding space as hard negatives (i.e., continuous sampling strategy).

\paragraph{Discrete Sampling Strategy.} \textit{Discrete} means the generator just samples discrete indexes of existing samples instead of generating a vector with continuous values \cite{PURE} like traditional GANs.
As one of the first attempts to apply GAN in information retrieval, the generator in IRGAN \cite{IRGAN} employs policy gradient based reinforcement learning to perform discrete negative sampling for confusing the discriminator, providing a direction for the subsequent applications such as GraphGAN \cite{GraphGAN} and NMRN \cite{NMRN}. Similar methods can also be applied to KGE tasks (e.g., KBGAN \cite{KBGAN}), where the generator approximates the score distribution of candidates to provide discrete negative triples, and the KG embedding model measures the quality of sampled data. With the help of high-order connectivity in the knowledge graph (KG), KGPolicy \cite{KGPolicy} works as a discrete sampler and adaptively receives knowledge-aware rewards to provide high-quality samples for the target RS model. 

\paragraph{Continuous Sampling Strategy.} Limited by finite samples, the advantage of adversarial training to mine high-quality negatives cannot be fully exploited by discrete sampling strategy, while another key research direction is to generate or synthesize negative samples from the continuous space. Incorporating easy negatives that are often overlooked, DAML \cite{DAML} generates synthetic negative samples adversarial to the learned metric with regularization, which taps the potential of adversarial methods to mine hard negatives in the feature space. Based on top of adversarially sampled negative samples, AdvIR \cite{AdvIR} generates informational negatives by adding adversarial perturbation to them. Not confined to the original data, HeGAN \cite{HeGAN} further leverages a generalized generator to directly sample latent nodes from a continuous distribution on heterogeneous information networks. In an adversarial end-to-end manner, AdCo \cite{AdCo} presents that challenging negative samples can also be adversarially generated together with the representation network.

\paragraph{Pros and cons.} {DNS methods in Section~\ref{sec:DNS} can be regarded as some specific forms of GAN-based NS, with query-aware scorers or positive sample-aware scorers serving as generators. By formulating the negative sampler and the contrastive learner as two different models playing a minimax game, GAN-based NS creates more room for exploring and exploiting high-quality negative samples, in a fully self-adjusted manner, rather than constrained by some fixed patterns set by heuristics. However, the additional generator module inevitably causes extra complexity in computational cost. Besides, the training process of minimax games are often not stable, so the final ideal equilibrium status can not always be guaranteed. To make the generator and} discriminator play games with each other instead of unilaterally training, the parameters may need pre-training \cite{KBGAN}, which further limits efficiency of the model. Meanwhile, adversarial negative sampling also suffers from false negative problems like DNS \cite{KGPolicy}. Solving the inherent problems of instability and inefficiency for adversarial sampling may lead to a long-awaited breakthrough.

\subsection{Efficient Negative Sampling}
Besides accuracy, another frequently desired characteristic of NS methods is the efficiency. In many real-world applications, such as pre-training large models in NLP or large-scale graph embedding, we cannot simply assume that the resource is unlimited and the full scope of candidate samples can be freely accessed. How to conduct efficient NS become an important and practical question. In this thread, related works can be summarized into two groups.

\paragraph{In-Batch Negative Sampling.}   An effective way to avoid resource reallocation is to reuse the samples within the same mini-batch for negative candidates, such as PBG \cite{PBG} in large-scale graph embeddings  and SimCLR \cite{SimCLR} in visual representation learning.
In this case, strategies can still be utilized to select hard negative samples, such as stratified sampling and negative sharing for collaborative filtering \cite{on-sampling-strategies-2017}, and adaptive batch scheduling (ABS) \cite{ABS} for dual encoders. In a distributed environment, a parallel negative sampling is proposed by GraphVite \cite{Graphvite} to cooperate multiple GPUs, which can also be regarded as the in-batch negative sampling with orthogonal sample blocks. Since batches are generated randomly, plain in-batch negative sampling is also a variant of RNS considering efficiency but the candidate set is limited by batch sizes.  
 
\paragraph{Caching Mechanism.} 
Another widely used method is to maintain a cache, such as a memory bank or queue, to store negative samples with asynchronous update mechanism, so that the in-batch sampling restriction is eliminated and more negative candidates can be involved during training. 
In knowledge graph embedding, NSCaching \cite{NSCaching} maintains a cache for the head and tail entity respectively, and randomly selects negative samples from the cache at each time. In each epoch, NSCaching updates the memory bank with the model's current predictive scores as distribution probability. However, refreshing all samples' probabilities is still time-consuming. Momentum Contrast (MoCo) \cite{MoCo} further stores the calculated representations in a queue following the rules of first-in-first-out (FIFO). Taking dynamic alteration into consideration, just updating the oldest representations is not enough. Therefore, there is still a trade-off between efficiency and effectiveness.

\section{Conclusion and Open Research Directions}

This paper presents an overview of negative sampling methods for contrastive representation learning, with a brief summary in Table~\ref{table1}. We generalize the task across various fields and systematically divide the existing methods into four categories. Meanwhile, we discuss the advantages and limitations for each category of method for better comparisons.

Taking into account the demonstrated challenges as well as the research progress in negative sampling, we summarize the following open research directions to advance this field.

\paragraph{Alleviating False Negatives.} As mentioned in Section~\ref{sec:DNS}, the so-called false negatives refer to the negative samples that will be positive in the future, which poses challenges for future research on negative sampling approaches. Although it is impossible to completely eliminate false negatives, some strategies can be employed to alleviate this problem beyond experimental settings. With a decomposition of true negative distribution, Chuang \textit{et al}. \shortcite{debiased-contrastive-learning} propose a contrastive learning goal to mitigate the negative sampling bias with a decomposition of true negative distribution. The method is empirically effective but limited to self-supervised learning. On the whole, the generalized method to eliminate false negatives is still an open problem to be solved.




\paragraph {Integrating Both Easy and Hard Negatives.} Though hard negative samples are critical in model training, it is inappropriate not to consider easy negative samples at all. Numerous easy negatives can not only exploited for generating hard negatives \cite{DAML}, but also maintain stability in the early training stage. Along with the training process of the target model, incorporating both the easy and hard negatives at multiple hard levels may lead to a more fine-grained representation. CuCo \cite{CuCo} integrates the concept of curriculum learning into negative sampling, which sorts the candidates based on the scoring function and learns from the easy to the difficult successively. In more fields considering negative sampling, arranging hardness of negative samples for better model optimization is worth exploring. 

\paragraph{Trade-off between Quality and Quantity.} Negative sampling approaches are mainly utilized to improve the quality of negative samples, while the sampling ratio between the positive samples and the negative ones determines the quantity. SimpleX \cite{SimpleX} proves that the most basic collaborative filtering method can be better than the state-of-the-art recommendation algorithm with the help of appropriate negative sampling ratio and loss function. Based on empirical evidence, more careful selection of the number and hardness of negative samples can both benefit CRL, while more or harder negative samples often imply lower efficiency. To some extent, the quality and quantity of negative samples can complement each other, which raises a question of balancing between the two. While there lacks a systematic study to realize the dynamic adjustment, a quality versus quantity trade-off is also a direction that needs to be fully explored. 

\paragraph{Without Negative Sampling.} Considering unresolved trade-offs involved in negative sampling, another direction is to obviate the need for negative sampling with efficient non-sampling strategy or even without negative samples. On the one hand, several methods (e.g., NS-KGE \cite{NS-KGE}) consider the whole data set as negative samples with dynamic weights with a controllable time complexity. On the other hand, without explicitly employing negative pairs over iterations, BYOL \cite{BYOL} discards negative samples in self-supervised learning and imposes additional regularization to avoid collapse of the representation, which achieves an even better result than negative-dependent methods such as SimCLR \cite{SimCLR}. 
Therefore, an exciting and promising research direction is to explore effective ways that can get rid of explicit negative sampling for CRL.

\bibliographystyle{named}
\bibliography{ijcai22}

\end{document}